\documentstyle[preprint,aps,epsfig]{revtex}
\begin{document}

\title{Soft Neutron Production in DIS:
 a Window to the Final State Interactions}
\author{M.~Strikman$^{a,b}$, M.G.~Tverskoy$^{b}$, M.B.Zhalov$^{b}$}
\address{ $^{(*)}$ Pennsylvania State University, University Park, PA, USA \\
 $^{(**)}$ S.Petersburg Nuclear Physics Institute, Russia \\}

\maketitle

\begin{abstract}
Recently E-665 reported the first measurement of  soft
 ($E_n \le 10 $MeV)
neutron production in deep inelastic scattering (DIS) off nuclei.
We report the first theoretical analysis of the data. We find that
the observed cross section can be quantitatively explained 
 as due to the final state interactions (FSI) of low energy
nucleons  ($E_N \sim 200-400 MeV$) produced in the 
elementary $\mu N$ interactions.
We argue that the data indicate strong a suppression of the  FSI's
of fast partons (hadrons) in DIS at high energies, and  
 that studies of the soft neutron
production would provide a new sensitive probe of
the dynamics of FSI's in
DIS both at fixed 
target energies and at the HERA $eA$ collider. 
\end{abstract}

PACS numbers:25.30Mr

\newpage

The final state interactions (FSI) in deep-inelastic scattering
 (DIS)  off nuclei
are known to provide a unique way of probing
 the space-time evolution of strong interactions. However, 
observing these effects is  challenging, especially if one wants to
study their energy dependence.
A number of experimental studies have
 focused on the 
production of 
leading hadrons in the current fragmentation region, see
\cite{e665h,busza,HERAstenhoven} and 
references  therein. The data indicate that
at high energies the absorption of the leading hadrons is very small
and hence most of the FSI occur at smaller rapidities.
Indications of such FSI were reported by  E665 in \cite{e665xe}.
This poses a serious problem for the planned studies of  FSI's at the
HERA collider in its eA mode of operation 
(for review of the project see the  report of the study group \cite{HERA})
since the present detectors are not sensitive to 
hadrons produced in the proton (nucleus)
fragmentation region. 
Hence we argued in \cite{HERA} that a study of soft neutrons
may provide one of very few windows at HERA for studies of FSI's.
Indeed such neutrons have energies  $\approx E_A/A$
 ($E_A$ is the energy of the nucleus) and very small
transverse momenta. As a result the H1 and ZEUS
 neutron detectors have nearly
100\% acceptance for  these neutrons \cite{CK}.  

The  production of soft neutrons with energies 
$E_n \le 10 MeV$
in DIS off nuclei appears to be 
 one of the very few observables which can be studied
both at fixed target energies using standard detectors of low energy neutrons
and at a $eA$ collider
 using a forward neutron calorimeter \cite{HERA,STZ}. 
The key question which we address here is whether soft neutrons
are sensitive enough to the FSI's in DIS 
to be of use for such studies.

The mechanism of the soft neutron production at
intermediate energies in pA scattering
(kinetic energies, $T_{inc} \le 0.5 - 1 $ GeV)
 is reasonably well understood -
such neutrons   are produced in the preevaporation and
evaporation of the residual nuclear system
left after interaction of the projectile with the target, see e.g.
Refs. \cite{codes}. For a recent review see 
\cite{BT}.
In the case of heavy nuclei these neutrons provide
the major channel of "cooling" of the residual system.

In the first approximation the total neutron multiplicity is
 proportional
to the number of nucleons knocked out from the nucleus
in  the initial (fast) stage of the  interaction.

Recently the first measurement of the soft neutron yields
in DIS was performed
by the
 E-665 collaboration \cite{e665n} at  FERMILAB. Neutrons with energies
 $E_N \le $ 10 MeV
were detected in high energy muon scattering off 
 a number of nuclear targets: D, C, Ca, Pb. A relatively
 small
{\bf average}
multiplicity of such neutrons, $\left<N_n(A)\right>$ was observed.
In particular, for the lead target  where the
data are most accurate
\begin{equation}
\left<N_n(Pb)(E_n \le  \mbox{10 MeV})\right>= 5 \pm 1.
\end{equation}

As far as we know  the only high energy data available in the high-energy data
bases 
 on the  production
of soft neutrons come from the ITEP (Moscow) experiment
 \cite{ITEP} for  incident
proton momenta of   1.4 $\le p_p^{inc} \le$ 8.5 GeV/c. We find that
the shapes of the spectra for
the overlapping energy range: 7.5 MeV $\le  E_n \le$ 10 MeV - 
are similar.
However,
the neutron multiplicity is much higher for the proton projectiles
(about a factor of 5-6 larger for $p-Pb$ collisions at $E_p \sim 10 GeV$)
which reflects a much larger number of wounded nucleons in $pA$ interactions.
In order to compare $pA$ and $\mu A$ data, we calculated 
 the number of neutrons produced per wounded
nucleon
$ {1\over A \sigma_{tot}(aN)}{d \sigma^3(a+A \rightarrow n +X) \over
d^3p/E}$, 
since in
 Glauber type models without secondary interactions
this quantity does not depend on the projectile \cite{FS81}.  
This quantity is 
still
 about a factor of two larger 
for the proton case, indicating that secondary interactions are more
 important for the proton projectile.

Average $x$ for the trigger used in \cite{e665n} is $\approx 0.015$.
For such $x$ nuclear shadowing for $Pb$ is very small:
 $\sigma_{tot}=\sigma_{imp.appr}-\sigma_{shad}$
with - $\eta \equiv \sigma_{shad}/\sigma_{imp.appr.}\approx 5\%$.   Using the 
Gribov theory of nuclear shadowing  which expresses
the shadowing effect, $\sigma_{shad}$ through
cross section of coherent diffraction in $\gamma^*N$ scattering,
 and the Abramovskii-Gribov-Kancheli
 cutting rules
 one finds that nuclear  shadowing leads to an increase of the cross section
of the diffractive events by $\sigma_{shad}$; the decrease of the 
 cross section of single interactions  from 
$\sigma_{imp.appr.}$ to $\sigma_{imp.appr.}-4\sigma_{shad}$, 
and emergence of the cross section of simultaneous  interactions with two nucleons
 equal to $2\sigma_{shad}$ , see \cite{FSAGK} for the recent discussion.
Taking into account that the number of soft neutrons
generated by the mechanism we discussed increases 
approximately by a factor of two for the 
events where two nucleons are knocked out, we find that the shadowing
effects lead to a slight increase of
average neutron multiplicity 
by a factor
 $\left(1-\eta\right)^{-1} \approx 1.05$
 for the kinematics of \cite{e665n}.
due to the shadowing effects. 
Hence, we will neglect this small effect in the further analysis.

It is generally assumed that in DIS high-energy hadrons are formed
at distances
\begin{equation}
l_{formation} \sim {p_h \over \mu^2}
\end{equation}
from the interaction point (see e.g. \cite{busza}), in which $p_h$ is 
the hadron momentum and the scale
  $\mu^2$ is $ \le 0.5$ GeV$^2$.  Hence at momenta above 10 GeV they
are formed 
beyond the nucleus
and  only  hadrons with energies $\le $ few GeV are
 involved in  FSI's with the residual nucleus.
 This picture  is consistent with a very weak $A$-dependence
of the leading particle spectra produced in DIS
at high energies, see e.g. \cite{HERAstenhoven,e665h}.
 The
most conservative assumption seems to be that only
slow recoiling nucleons
produced in the elementary lepton-nucleon DIS 
in the target fragmentation region
reinteract
with  the nucleus. The spectrum of the recoil nucleons can be approximated at small
$x$ and not very large energies, where the  triple Pomeron
contribution is still small,
as
\begin{equation}
{1 \over \sigma_{\gamma^*p}}{d\sigma^{\gamma^*+p \rightarrow N+X}
\over dz d^2p_t}
\propto {\exp(-ap_t^2) \over {\sqrt{z}}}
\label{eq3}
\end{equation}
with $a^{-1/2}=\left<p^2_t\right>^{1/2} \sim 0.4$GeV/c,
 see discussion in \cite{FS81}. Eq.\ref{eq3} is consistent with 
 the limited data of the BEBC neutrino experiment \cite{BEBC}.
Here $z$ is Feynman $x$ for the nucleons.
 Typical kinetic energies of the nucleons produced 
 in the nucleus rest frame
are $\sim $ 200-400 MeV,
and therefore by eq.2 they are formed within a modest nucleus radius.
 Soft neutrons should be produced
both in 
the decay of the hole  formed by the removal of 
a nucleon in the
elementary lepton-nucleon  DIS,   
and  in  the reinteractions of the secondary slow nucleons
originating from the elementary DIS process.

Contribution of these two effects
should be considered
as {\it a lower limit } for the rate of the soft neutron  production.
This is because   there exist  other processes
for which the standard time formation
arguments  do not hold as well.  For example  the slow 
pion absorption process is known to 
increase the excitation energy of a nucleus by
60 -- 70 MeV and  yield, on
 average, several
 additional
soft neutrons
per event, see e.g.
 \cite{BT}. Contributions of of high-energy parton rescatterings,
formation of hadrons inside the 
nuclei would further increase the neutron yield.

We focus on the case of the DIS scattering off lead, since the muon data
\cite{e665n} are much more accurate in this case.
To calculate the rate of the soft neutron production we considered the
following model of the  reaction:
(i) a nucleon is  removed from
any point in the nucleus with
a probability proportional to the nuclear density; (ii) an energy $W_n$
is assigned to  the produced nucleon based on cross section of the
 process $\mu +N \to \mu'+N'+X$ as given by Eq.(3); the interactions of 
the  nucleon produced in DIS are modeled
with a  Monte Carlo code for
hadron -nucleus interaction which  describes   all stages of
the process: the cascade, which includes  knock-out of secondary
nucleons,
 production and subsequent interaction of pions;
preevaporation; and evaporation of neutrons and charged particles.
This  Monte Carlo code is very close to the  codes are used over the years
 to describe cascades in $pA$ scattering
at $E_N \le 1 GeV$,  see e.g. \cite{codes,BT}.
Note that in the original experimental paper \cite{e665n}
 only a qualitative analysis of the data was presented, assuming 
that neutrons originate from the decay of the hole formed
by removal of a nucleon in the DIS $\mu N$
interaction. Production
of the soft neutrons  in the FSI's of the
nucleons produced in the primary DIS $\mu N$
scattering was not considered though in our
calculation these FSI's provide most of the soft neutrons.

The code was tested  using available $pA$ data at intermediate 
energies \cite{ITEP,n113,n585}.
We found that the code produces a good description of the neutron yields
in the kinematics of interest, see Fig.1,2. 
The curves are about 20\% above the data which is consistent with the typical
20\% accuracy of such codes. 
This may also reflect uncertainties in the normalization of the data.

We find that the calculated  spectrum of the
soft  neutrons is  consistent with the E-665 data as reported in 
the erratum \cite{e665n},  see solid line in Fig.3.

We want to emphasize that the calculated yield of neutrons 
depends weakly  on the model used for the spectrum of nucleons produced 
in the elementary reaction. In particular, assuming that all nucleons
are produced with energy of 200 or 400 MeV (dashed and dotted curves in Fig.3)
practically do not change our result. To illustrate further 
our weak sensitivity to the model of the nucleon production 
we present in Fig.4
the multiplicity of the produced neutrons for different cutoffs in
$E_n $. One can see that it  weakly depends on the 
kinetic energy of the generated nucleon, $W_n$,
 for the kinetic energies of interest:
200 MeV $\le W_n \le 500$ MeV. Note here that 
$\left<W_n \right>$ for the model corresponding to Eq.\ref{eq3}
is 
$\approx 300 $ MeV.

We estimate
\begin{equation}
\left<N_n(Pb)(E_n \le  \mbox{10 MeV})\right>_{lower~~ limit}
= 6 \pm 1.5,
\label{MC}
\end{equation}
which  is reasonably close to the experimental number of $5 \pm 1$.

It is worth noting that there is a trend in the E665-data  for
$\left<N_n(Pb)(E_n \le  \mbox{10 MeV})\right>$
to fall with increasing  $q_0 \equiv 
\nu$ (see Fig.2 of \cite{e665n}). This may be  due to two causes:
(i) the  decrease of the probability for the leading hadrons to reinteract with
 nuclei with increasing $\nu$ as observed in \cite{e665h}, and 
(ii) a softening of the spectrum of the nucleons produced
in the elementary process with decreasing  $x$ (larger
 $\nu$ in the data sample of E665 correspond to smaller  average $x$).

 The multiplicity obtained in our ``minimal'' 
 model of soft
neutron production  seems to leave very little
room for the further processes of FSI of fast hadrons in the nuclei.
The only alternative we could think of is that suppression of
the FSI of produced nucleons starts at much lower energies than it is
usually thought, say 1-2 GeV. 

To illustrate the  sensitivity of the
E-665 to a model of the FSI
let us estimate 
 $\left<N_n(A)\right>$ 
 in a class  of the models where it is assumed that the 
leading quark can interact with effective cross section
$\sigma_{eff}$ of 10-20 mb, see e.g. discussion in \cite{HERA,bialas}.
The number of extra interactions due to this mechanism can be estimated as
\begin{equation}
\delta(A)={1\over 2}{A-1\over A^2}\sigma_{eff}\int d^2b T^2(b),
\end{equation}
where $T(b)$ is the standard thickness function
 $T(b)=\int dz \rho_A(b,z)$.
In the case of lead  target this leads to 
$\approx 0.75
\sigma_{eff}$ additional  interactions per event where $\sigma_{eff}$  
is measured in fm$^2$. Low energy  nucleons
produced in these interactions would have
the energy distribution close to the one
in the elementary reaction and hence generate 
soft neutrons with the same efficiency as the 
``minimal'' mechanism.
 Thus $\sigma_{eff} =10 (20) mb$
would lead to 
the estimated soft neutron  multiplicity of 8.75$\pm$ 1.75 (12.5 $\pm$ 2.5)
(to produce a conservative estimate 
we include here the $pA$ data based adjustment factor of 0.8
for the theoretical value
of eq.\ref{MC}
which is well above  the experimental value of 5$\pm 1$.
We are planning to combine our MC code with a number of MC codes
 for production of high energy particles in $eA$ scattering to constrain parameters of these models using E665 data.

It would be really challenging to reconcile a low neutron multiplicity
observed by E665 with the indications of a noticeable FSI's in production of  
 faster particles in the
 nuclear fragmentation region obtained in the same experiment,
 see discussion in \cite{venus}. This may require a more complicated dynamics of FSI's than currently  envisioned. It also 
calls for a systematic 
experimental study of the energy dependence of
the neutron 
yield starting from the incident energies available in the HERMES
experiment 
and in the forthcoming COMPASS experiment.
One can expect a gradual decrease of the rate of soft
 neutron production
with increasing of  $\nu$ and $ Q^2$.  It would be interesting
to study correlations between the neutron multiplicity
and the spectrum of leading hadrons ($z$-distribution, $p_t$
broadening, etc). It is necessary also
 to repeat the  E665 experiment at higher energies
to check rather amazing finding of this experiment of the low rate of
 production of soft neutrons.
If the decrease with $E_{inc}$ and low neutron  multiplicity at
$E_{\mu} \ge$ 200 GeV are confirmed,
the soft neutrons would provide a perfect tool
to look for relatively rare final
 state interactions in DIS at the  HERA collider
at small $x$ by selecting events 
with much larger than average neutron multiplicity.
For example it would be possible to investigate the $p_t$ broadening effects
\cite{sterman,yura} via a study of the dependence of
$\left<N_n\right>$ on the transverse momentum of the leading hadron
in the current fragmentation region.

At the same time there are a few questions which could be
addressed at the intermediate energy lepton facility, like
TJNAF. The first question concerns  the estimate  of the contribution
to the  soft neutron multiplicity from the  decay of the hole state
produced by removing  one nucleon in DIS. This problem is tightly
connected with the study of the one-hole nuclear spectral function
which is the subject of a number of  $(e,e^\prime p)$ experiments
that  are under way  at TJNAF.
It would  be of interest to measure the spectrum 
of neutrons produced in the decay of the hole states, and
its dependence on the missing energy. Hall C may provide
possibilities for such study using a neutron detector 
in a combination with a high resolution $(e,e'p)$ experiment.
Such measurements also may be performed as 
 a part of the recently approved
$(e,e^\prime pn)$ experiment \cite{newexp}.
 The
experiment of the
same kind but measuring yield of soft neutrons as a function of the energy
transferred by electron to nuclear proton (low energy resolution)
can help to obtain the reliable estimate of the soft neutron multiplicity
due to the FSI's of the intermediate energy protons
with medium and heavy mass nuclei in lepton-nucleus interaction.

We would like to thank K.Griffioen numerous suggestions and 
providing details of the E665 kinematics.
We also thank  L.Frankfurt, A.E.L.Dieperink
and M.Sargsian for discussions.

The research was supported by the DOE grant DE-FG02-93ER40771
and the ISF grant
SAK 000.

\newpage

\vspace{-10cm}
\begin{figure}
\centerline{
\epsfig{file=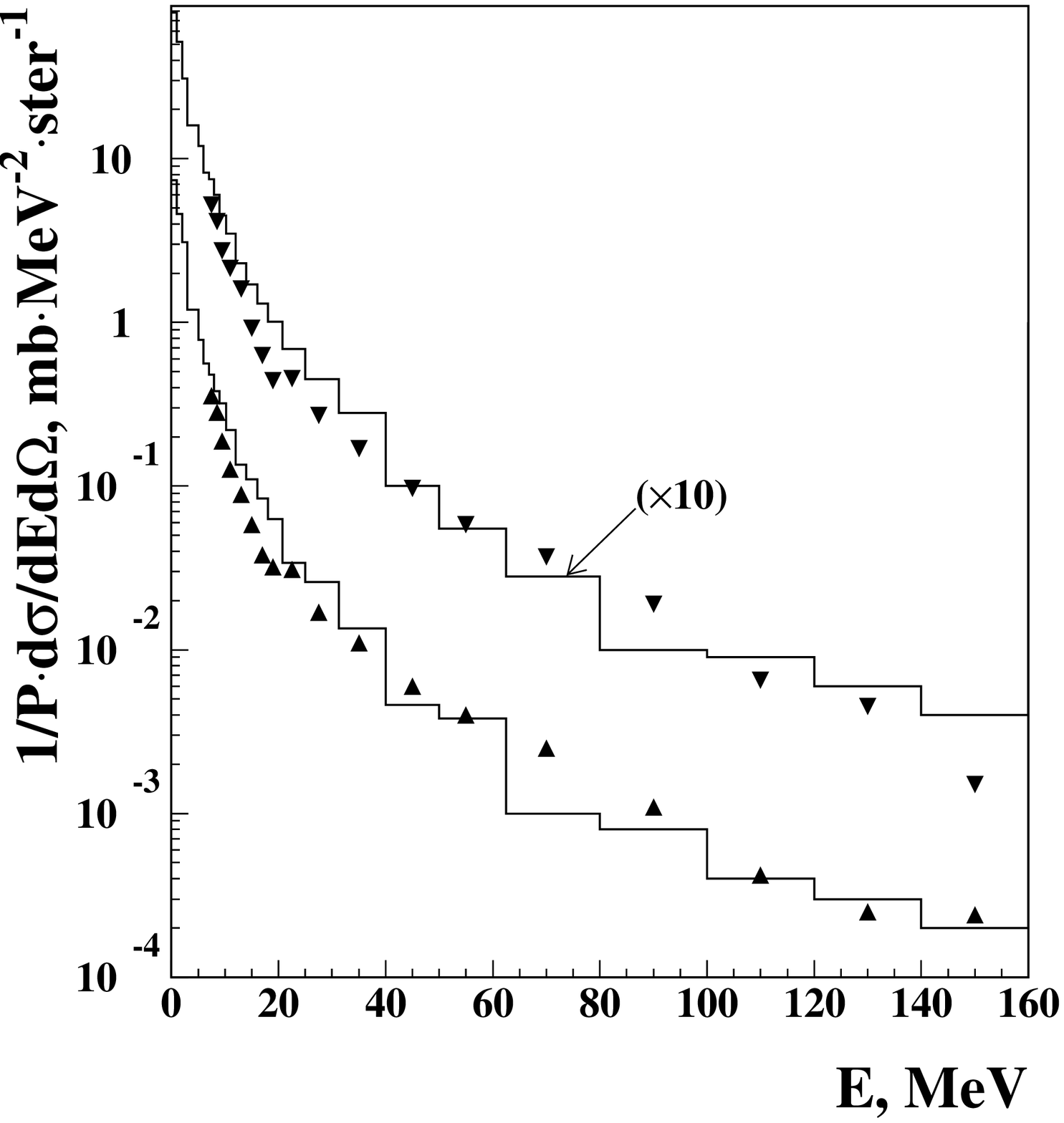,width=15cm,height=15cm}}
%\vspace{-4.5cm}
\caption{Comparison of the results of the Monte Carlo cascade-evaporation
 calculation of  the neutron
spectra in \mbox{$p+Pb \rightarrow n +X$} process 
with the ITEP  data \protect\cite{ITEP} at $P_p=1.4 GeV/c$ (lower curve)
and $P_p=2 GeV/c$ (upper curve) for $\theta_n=120^o$.
The overall experimental errors which
 are not shown in the figure are
 $\sim 20\%$ }
\end{figure}

\newpage 
\vspace{-10cm}
\begin{figure}
\centerline{
\epsfig{file=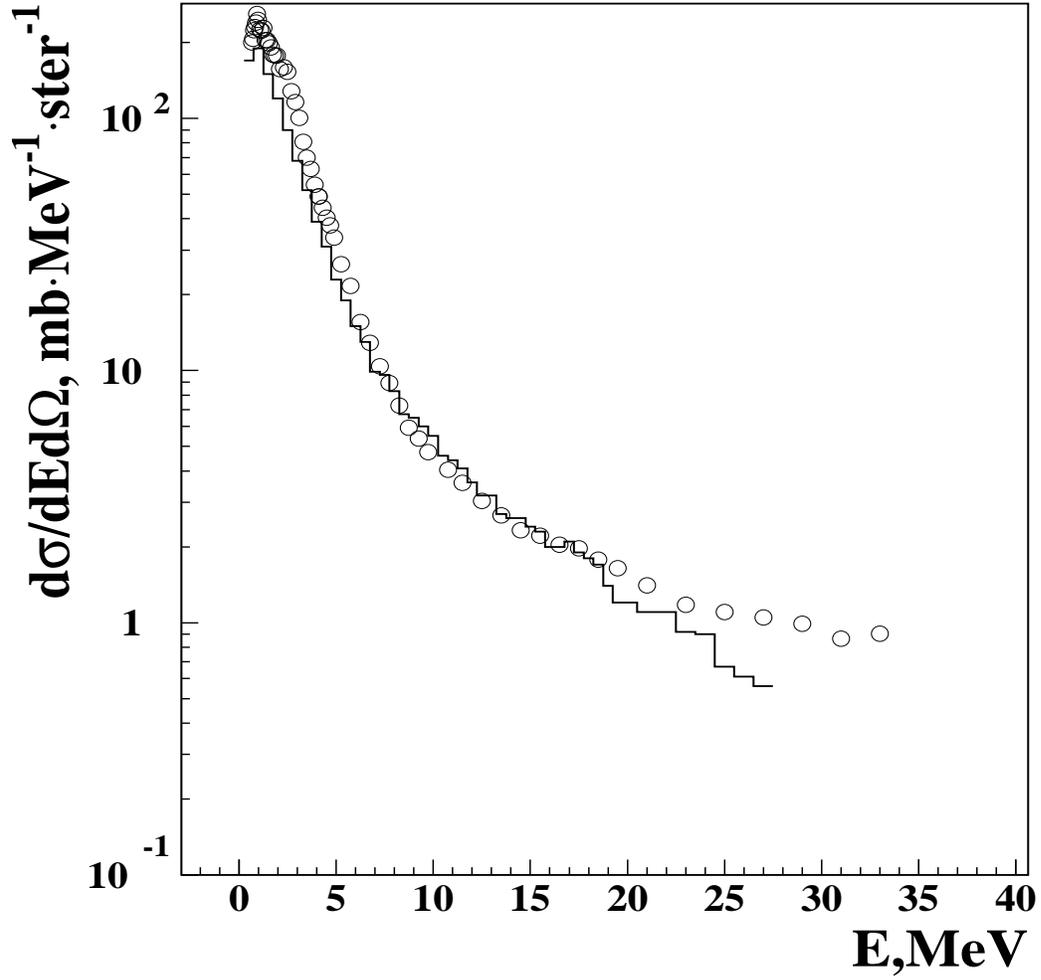,width=15cm,height=15cm}}
%\vspace{-4.5cm}
\caption{Comparison of the results of the Monte Carlo cascade-evaporation
 calculation of  the neutron
spectra in \mbox{$p+Pb \rightarrow n +X$} process at $E_p^{inc}$=113 MeV
with the data \protect\cite{n113}.}
\end{figure}

\vspace{-10cm}
\begin{figure}
\centerline{
\epsfig{file=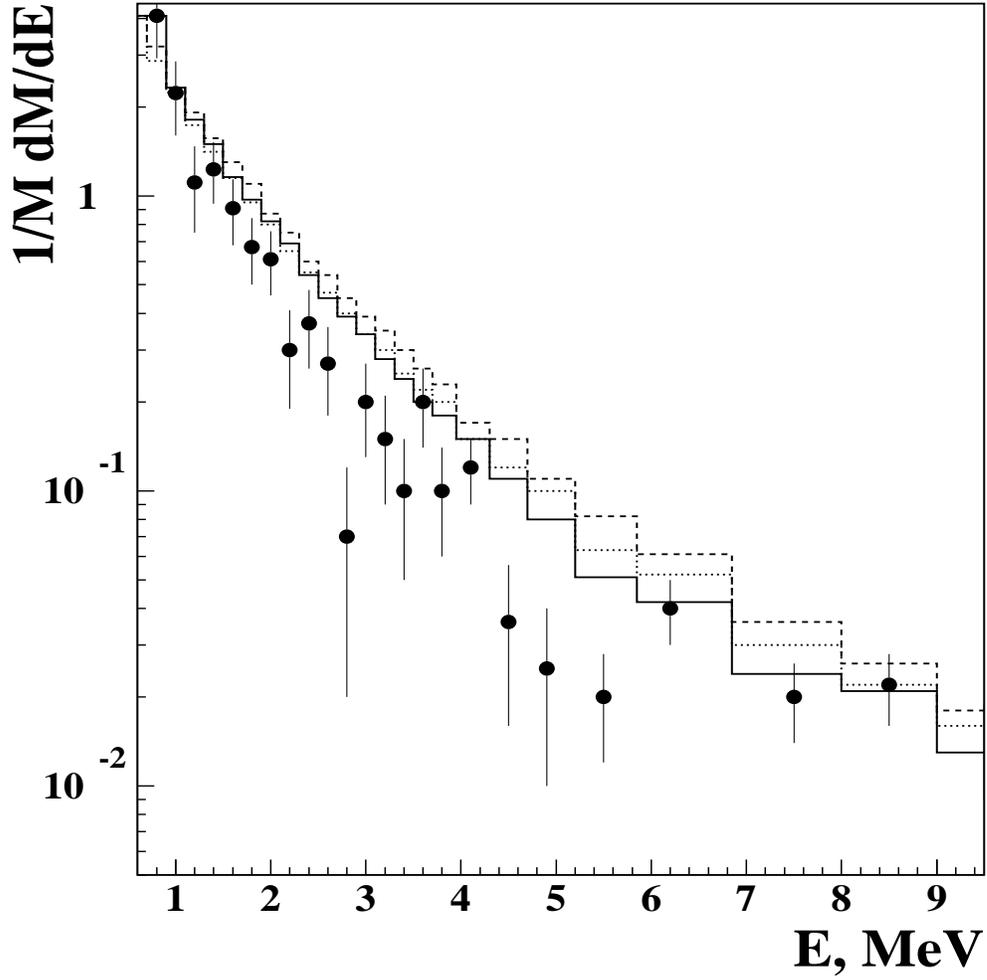,width=15cm,height=15cm}}
\vspace{3.5cm}
\caption{Comparison of the results of the Monte Carlo cascade-evaporation
 calculation of  the neutron
spectra in \mbox{$\mu+Pb \rightarrow n +X$} process  
with E-665 data \protect\cite{e665n}.}

\end{figure}

\newpage
\begin{figure}
\centerline{
\epsfig{file=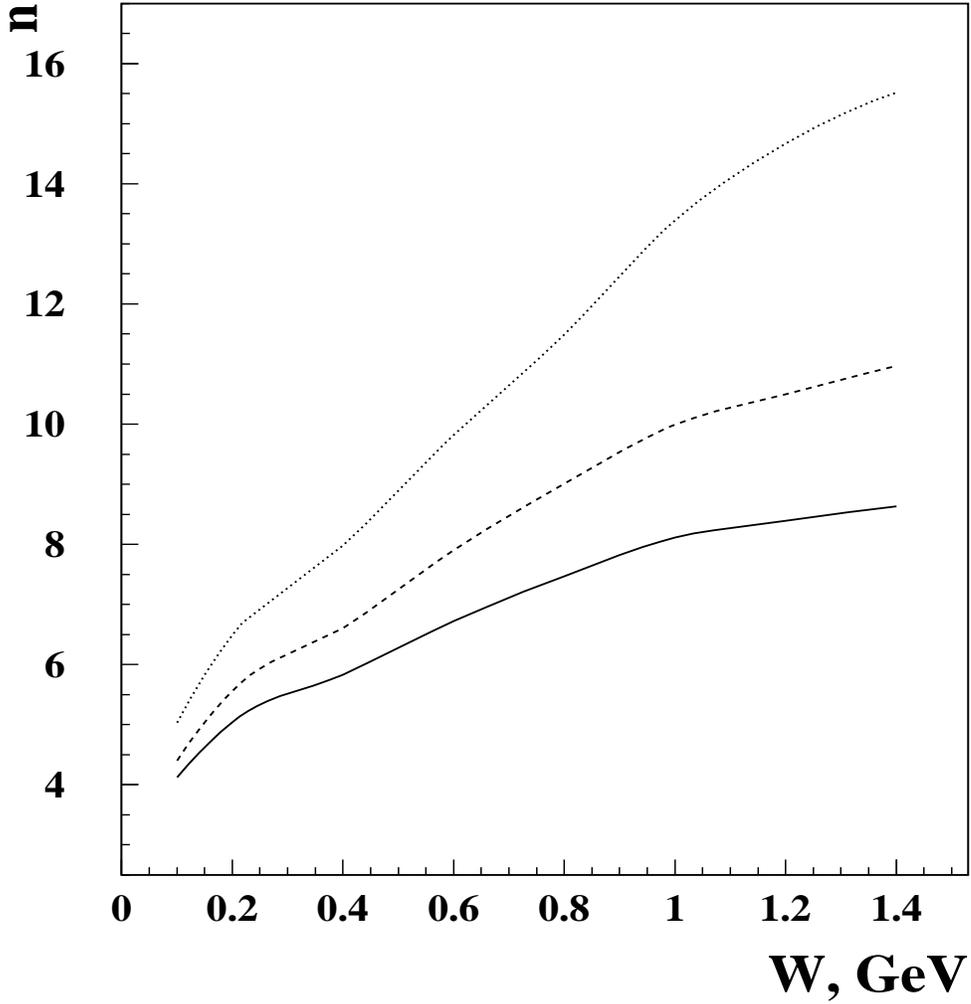,width=15cm,height=15cm}}
%\vspace{-2.5cm}
\caption{Multiplicity of neutrons produced in the process where a nucleon with
energy $W$ was produced in $Pb$ for different energy intervals of neutron
energy. Dotted, solid,
dashed curves are multiplicities of evaporated nucleons with
kinetic energies $T_n$
for $0 \le T_n \le 6$ MeV,  $0 \le T_n \le 10$ MeV,
$0 \le T_n \le 50$ MeV; dashed-dotted curve is the  total nucleon
multiplicity.}
\end{figure}

\end{document}